\documentclass[aps,superscriptaddress,twocolumn,10pt,nofootinbib,floatfix]{revtex4-1}

\usepackage{amssymb}
\usepackage{amsmath}
\usepackage{times}
\usepackage{graphicx}
\usepackage{color}
\usepackage{bm}
\usepackage{dcolumn}
\usepackage{appendix}
\usepackage[colorlinks=true, letterpaper=true, pdfstartview=FitV, linkcolor=blue, citecolor=blue, urlcolor=blue]{hyperref}
\usepackage[normalem]{ulem}

\setcitestyle{square}

\makeatletter
\renewcommand\@biblabel[1]{#1.}
\makeatother

\begin{document}
	
	\title{Electrically Controllable Crystal Chirality Magneto-Optical Effects in Collinear Antiferromagnets}
	
	\author{Xiaodong Zhou}
	\affiliation {Key Laboratory of Advanced Optoelectronic Quantum Architecture and Measurement, Ministry of Education, School of Physics, Beijing Institute of Technology, Beijing 100081, China}
	
	\author{Wanxiang Feng}
	\email{wxfeng@bit.edu.cn}
	\affiliation {Key Laboratory of Advanced Optoelectronic Quantum Architecture and Measurement, Ministry of Education, School of Physics, Beijing Institute of Technology, Beijing 100081, China}
	
	\author{Xiuxian Yang}
	\affiliation {Key Laboratory of Advanced Optoelectronic Quantum Architecture and Measurement, Ministry of Education, School of Physics, Beijing Institute of Technology, Beijing 100081, China}
	
	\author{Guang-Yu Guo}
	\affiliation {Department of Physics and Center for Theoretical Physics, National Taiwan University, Taipei 10617, Taiwan}
	\affiliation {Physics Division, National Center for Theoretical Sciences, Hsinchu 30013, Taiwan}
	
	\author{Yugui Yao}
	\email{ygyao@bit.edu.cn}
	\affiliation {Key Laboratory of Advanced Optoelectronic Quantum Architecture and Measurement, Ministry of Education, School of Physics, Beijing Institute of Technology, Beijing 100081, China}
	
	\date{\today}

	\begin{abstract}
		The spin chirality, created by magnetic atoms, has been comprehensively understood to generate and control the magneto-optical effects.  In comparison, the role of the crystal chirality that relates to nonmagnetic atoms has received much less attention.  Here, we theoretically discover the crystal chirality magneto-optical (CCMO) effects, which depend on the chirality of crystal structures that originates from the rearrangement of nonmagnetic atoms.  We show that the CCMO effects exist in many collinear antiferromagnets, such as RuO$_{2}$ and CoNb$_{3}$S$_{6}$, which has a local and global crystal chirality, respectively.  The key character of the CCMO effects is the sign change if the crystal chirality reverses.  The magnitudes of the CCMO spectra can be effectively manipulated by reorienting the N{\'e}el vector with the help of an external electric field, and the spectral integrals are found to be proportional to magnetocrystalline anisotropy energy.
	\end{abstract}
	
	\maketitle

	Among large family members of the magneto-optical  (MO) effects, the Kerr~\cite{Kerr1877} and Faraday~\cite{Faraday1846} effects discovered in the mid-18th century are the representatives, which describe the rotations of the planes of polarization of the linearly polarized light reflecting from and transmitting through magnetic media, respectively.  After they were discovered over a century, a critical theoretical explanation of them was revealed on the basis of the band theory~\cite{Argyres1955}.  Since then, two basic understandings for the MO effects were broadly accepted~\cite{Reim1990,Mansuripur1995,Ebert1996,Antonov2004,Kuch2015}: The magnitudes of MO spectra are proportional to the spontaneous magnetization of magnetic media such that antiferromagnets (AFMs) with zero net magnetization are intuitively believed to present no MO signals; The simultaneous presence of band exchange splitting (due to finite net magnetization) and spin-orbit coupling is the sole physical origin of the MO effects.
	
	Nevertheless, new insights into the MO effects emerged recently.  On the one hand, the MO effects were surprisingly found in noncollinear~\cite{WX-Feng2015,Higo2018,Balk2019,MX-Wu2020,Wimmer2019,XD-Zhou2019a} or collinear~\cite{Sivadas2016,FR-Fan2017,K-Yang2020,De2020} AFMs even though their net magnetization is vanishing.  The MO Kerr effect was first predicted to exist in noncollinear (coplanar) AFMs Mn$_{3}X$ ($X=$ Rh, Ir, Pt)~\cite{WX-Feng2015} by the natural lack of a good symmetry $\mathcal{TS}$ ($\mathcal{T}$ is the time-reversal symmetry; $\mathcal{S}$ is a spatial symmetry) that is responsible for band exchange splitting.  The collinear AFMs (e.g., MnPSe$_{3}$) were then demonstrated to host the Kerr effect if an electric field is introduced to break the $\mathcal{TS}$ symmetry~\cite{Sivadas2016}.  On the other hand, the topological MO effects that originate from the scalar spin chirality was discovered in noncollinear (noncoplanar) AFMs~\cite{WX-Feng2020}.  The band exchange splitting and spin-orbit coupling are not indispensable for the topological MO effects and their quantization~\cite{WX-Feng2020}, in sharp contrast to the conventional MO effects.

	The above discoveries can be mainly ascribed to chiral spin textures created by magnetic atoms.  However, the role of nonmagnetic atoms that is related to crystal chirality has not been completely understood till now.  In this work, using the first-principles calculations and group theory analyses, we reveal a new class of MO effects in collinear AFMs, termed crystal chirality magneto-optical (CCMO) effects.  The MO Kerr and Faraday spectra change their signs when the crystal structure's chirality alters from the left-handed state to the right-handed state or vice versa.  Moreover, the magnitudes of MO spectra can be effectively manipulated by reorienting the N{\'e}el vector with the help of an external electric field.  The spectral integrals are found to be proportional to magnetocrystalline anisotropy energy (MAE), which has not been previously reported in AFMs.  The CCMO effects discovered here may be a promising optical means to simultaneously detect crystal structure's chirality and N{\'e}el vector's orientation.
	
	\begin{figure*}
		\includegraphics[width=2\columnwidth]{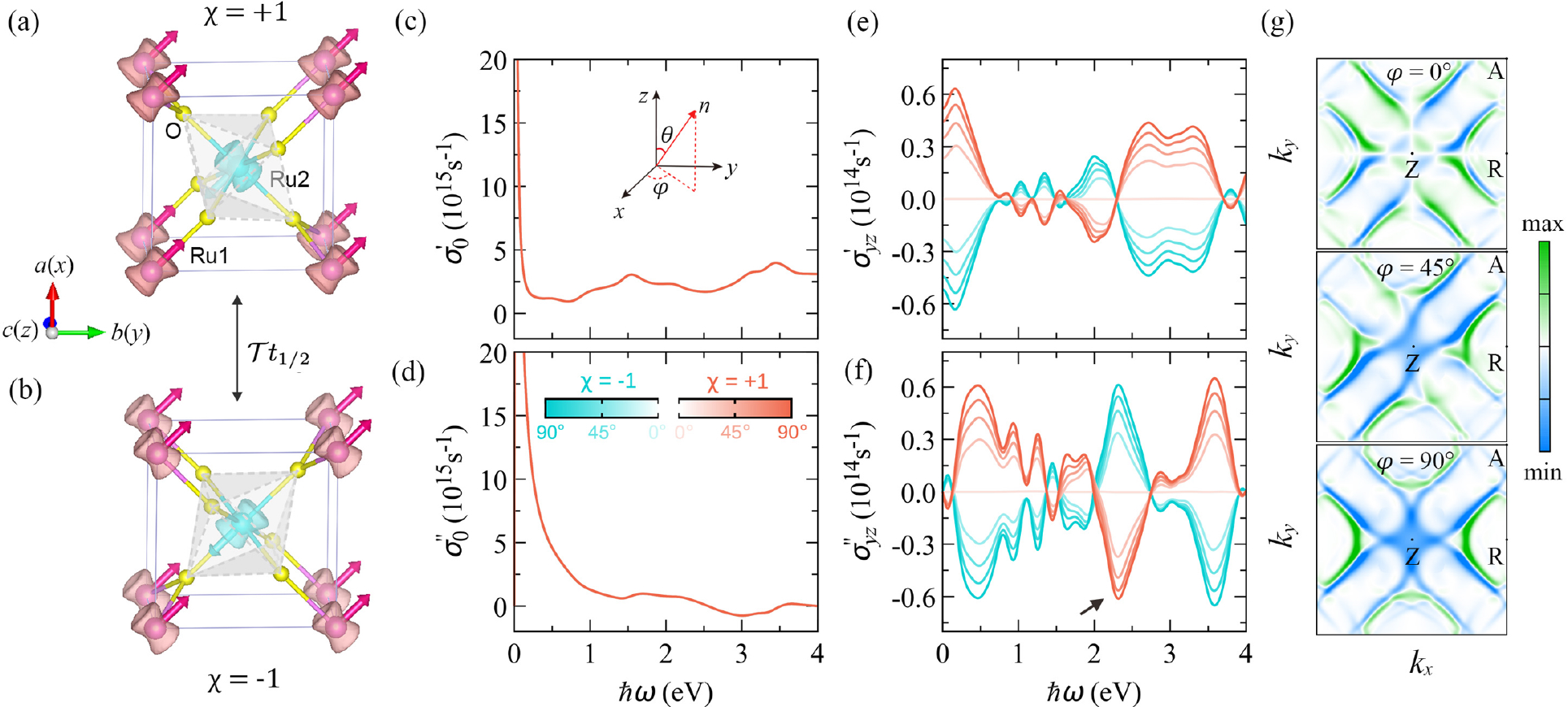}
		\caption{(a)(b) The crystal structures of RuO$_{2}$ with the right-handed ($\bm{\chi}=+1$) and left-handed ($\bm{\chi}=-1$) chiralities.  Pink and cyan spheres represent two magnetic Ru atoms (Ru1 and Ru2), whereas yellow spheres are nonmagnetic O atoms.  The dashed lines outline the octahedron formed by six O atoms.  The magnetic density isosurfaces surrounding two Ru atoms differ by 90 degrees.  Pink and cyan arrows label the spin orientations of two Ru atoms, and the N{\'e}el vector is schematically plotted along the [110] direction.  Two chiral structures are related to each other by the symmetry $\mathcal{T}t_{1/2}$.  (c)(d) Real and imaginary parts of the diagonal element and (e)(f) real and imaginary parts of the off-diagonal element of optical conductivity for $\bm{\chi}=\pm1$ states of RuO$_{2}$ when the N{\'e}el vector $\bm{n}$ rotates within the (001) plane ($\theta = 90^{\circ}$ and $0^\circ\leqslant\varphi\leqslant90^\circ$).  The inset in (c) shows the spherical coordinates ($\theta$, $\varphi$) used for describing the N{\'e}el vector $\bm{n}$.  (g) The momentum-resolved $\sigma^{\prime\prime}_{yz}$ for the $\bm{\chi}=+1$ state when $\varphi=0^{\circ}$, $45^{\circ}$, and $90^{\circ}$, at the incident photon energy of 2.31 eV that is marked by an arrow in (f).}
		\label{fig1}
	\end{figure*}

	The crystal structure's chirality is generated from different arrangements of nonmagnetic atoms by keeping the structure's space group unchanged.  The crystal chirality manifests in many magnetic materials with different dimensions, e.g., three-dimensional (3D) rutile RuO$_{2}$~\cite{Smejkal2020} and $M$F$_{2}$ ($M=$ Mn, Ni)~\cite{Wu2016,LD-Yuan2020,X-Li2019}, quasi-2D $N$Nb$_{3}$S$_{6}$ ($N=$ V, Cr, Mn, Fe, Co, Ni)~\cite{Anzenhofer1970,Parkin1980,Smejkal2020}, and 2D SrRuO$_{3}$ monolayer~\cite{Samanta2020}.  Here, we take RuO$_{2}$ and CoNb$_{3}$S$_{6}$ as two prototypes, which has a local and global crystal chirality, respectively.
	
	Rutile RuO$_{2}$ has long been considered to be a Pauli paramagnet~\cite{Ryden1970}, while several recent studies revealed a collinear antiferromagnetic order under room temperature~\cite{Berlijn2017, ZH-Zhu2019,ZX-Feng2020}.  RuO$_{2}$ is a centrosymmetric system (crystallographic space group $P4_{2}/mnm$) with inversion centers at magnetic Ru atoms, therefore, it can not be regarded as a usual chiral crystal~\cite{Flack2003,Nespolo2018}.  However, one can still define a \textit{local} crystal chirality for RuO$_{2}$~\cite{Smejkal2020}, given by $\bm{\chi} = \bm{d}_{\textrm{Ru1O}} \times \bm{d}_{\textrm{ORu2}}$ ($ \bm{d}_{\textrm{Ru1O}}$ and $\bm{d}_{\textrm{ORu2}}$ are two vectors connecting the path Ru1-O-Ru2), because nonmagnetic O atoms are not inversion centers.  Evidently, RuO$_{2}$ has two opposite chiral states, i.e., right-handed chirality $\bm{\chi}=+1$ [Fig.~\ref{fig1}(a)] and left-handed chirality $\bm{\chi}=-1$ [Fig.~\ref{fig1}(b)].  The two chiral states are energetically degenerate, but their magnetic densities are redistributed by rotating 90 degrees.  Interestingly, one chiral state is related to the other one via a combined symmetry $\mathcal{T}t_{1/2}$, where $t_{1/2}=[0.5,0.5,0.5]$ is the half-unit cell translation.  For each chiral state, the N{\'e}el vector, $\bm{n}=(\bm{n}_{\text{Ru1}}-\bm{n}_{\text{Ru2}})/2$ ($\bm{n}_{\text{Ru1}}$ and $\bm{n}_{\text{Ru2}}$ are the spins of two Ru atoms), can be reoriented by the current-induced spin-orbit field, like in CuMnAs~\cite{Wadley2016,Godinho2018}.  Our first-principles calculations~\cite{SuppMater} show that the N{\'e}el vector of RuO$_{2}$ points along the [001] direction with a MAE of 2.76 meV/Ru atom, which accords well with the previous theoretical results~\cite{Berlijn2017,Smejkal2020}.  While the resonant X-ray scattering data indicate that the N{\'e}el vector slightly cants from the [001] direction and has nonzero [100] and [010] components~\cite{ZH-Zhu2019}.  Such a generic magnetization direction suggests that the N{\'e}el vector can be easily tuned, giving rise to the coupling between the crystal chirality $\bm{\chi}$ and N{\'e}el order $\bm{n}$.
	
	The calculation of optical conductivity is the crucial step to obtain the MO Kerr and Faraday spectra~\cite{SuppMater}.  For convenience, the vector-form notation of optical Hall conductivity, $\bm{\sigma}=[\sigma_{yz},\sigma_{zx},\sigma_{xy}]=[\sigma^{x},\sigma^{y},\sigma^{z}]$, is used here.  Since $\bm{\sigma}$ can be regarded as a pseudovector, like the spin, its nonvanishing components can be identified by acting on each group element of relevant magnetic groups~\cite{XD-Zhou2019a}.  And the results of symmetry analyses for $\bm{\sigma}$ can be directly applied to the Kerr and Faraday angles, $\bm{\phi}_{K}=[\phi^{x}_{K},\phi^{y}_{K},\phi^{z}_{K}]$ and $\bm{\phi}_{F}=[\phi^{x}_{F},\phi^{y}_{F},\phi^{z}_{F}]$ (see Eqs. (\textcolor{blue}{S6}) and (\textcolor{blue}{S7}) in Ref.~\cite{SuppMater}).

	First, we prejudge the nonvanishing components of $\bm{\sigma}$  by considering the N{\'e}el vector $\bm{n}$ lying on the (100), (010), and (001) planes.  When $\bm{n}$ points along the [001] direction, RuO$_{2}$ has a magnetic space group $P4_{2^{\prime}}/mnm^{\prime}$, which contains two glide mirror planes $\mathcal{M}_{[100]}t_{1/2}$ and $\mathcal{M}_{[010]}t_{1/2}$.  The mirror $\mathcal{M}_{[100]}$ changes the signs of $\sigma^{y}$ and $\sigma^{z}$, but preserves $\sigma^{x}$; the half-unit cell translation $t_{1/2}$ plays a nothing role on $\bm{\sigma}$; thus, the combined operation $\mathcal{M}_{[100]}t_{1/2}$ restricts optical Hall conductivity to be a shape of $\bm{\sigma}=[\sigma^{x},0,0]$.  Similarly, another glide mirror plane $\mathcal{M}_{[010]}t_{1/2}$ gives rise to $\bm{\sigma}=[0,\sigma^{y},0]$. Therefore, under the group $P4_{2^{\prime}}/mnm^{\prime}$, the optical Hall conductivity is zero, $\bm{\sigma}=[0,0,0]$.  When $\bm{n}$ deviates away from the [001] direction and rotations within the (100) and (010) planes, the symmetries $\mathcal{M}_{[010]}t_{1/2}$ and $\mathcal{M}_{[100]}t_{1/2}$ are absent, giving rise to $[\sigma^{x},0,0]$ and $[0,\sigma^{y},0]$, respectively.  Within the (001) plane, the magnetic space group always contains the symmetry $\mathcal{TM}_{[001]}$, which changes the sign of $\sigma^{z}$ but preserves $\sigma^{x}$ and $\sigma^{y}$.  Therefore, $\sigma^{x}$ and $\sigma^{y}$ are potentially nonzero.  In particular, if $\bm{n}$ points along the [100] ([010]) direction, only $\sigma^{y}$ ($\sigma^{x}$) can be nonzero due to the role of $\mathcal{M}_{[010]}t_{1/2}$ ($\mathcal{M}_{[100]}t_{1/2}$).  To summarize, once $\bm{n}$ cants slightly from the [001] direction, the optical Hall conductivity and the MO spectra turn to be nonzero.

	From the symmetry point of view, the effect of crystal chirality on $\bm{\sigma}$ differs from that of spin chirality.  In noncollinear AFMs Mn$_{3}Z$N ($Z=$ Ga, Zn, Ag, Ni)~\cite{XD-Zhou2019a}, the reversal of spin chirality changes nonvanishing components and magnitudes of $\bm{\sigma}$.  While two crystal chirality states of RuO$_{2}$ are related by the symmetry $\mathcal{T}t_{1/2}$, only the sign of $\bm{\sigma}$ will be changed as $\bm{\sigma}$ is odd under $\mathcal{T}$.  Thus, the sign change can also be achieved by reversing the N{\'e}el vector $\bm{n}$.  In fact, the sign of $\bm{\sigma}$ for RuO$_{2}$ is determined by the sign of $\bm{n}\cdot\bm{\chi}$.

	Next, we discuss the optical conductivity calculated from the first-principles methods.  In light of a recently experimental fabrication of the [001]-orientated RuO$_{2}$ thin films with in-plane magnetization~\cite{ZX-Feng2020}, we pay our attention to the $\bm{n}$ lying within the (001) plane.  Figs.~\ref{fig1}(c) and~\ref{fig1}(d) display the diagonal element of optical conductivity, $\sigma_{0}=(\sigma_{yy}+\sigma_{zz})/2$, for two chiral states of RuO$_{2}$ as a function of $\varphi$.  The real part $\sigma_{0}^{\prime}$ measures the average in the absorption of left- and right-circularly polarized light.  It has two absorptive peaks at 1.5 eV and 3.5 eV and diverges in the low-frequency region due to the inclusion of Drude term~\cite{SuppMater}.  The imaginary part $\sigma_0^{\prime\prime}$ can be obtained from $\sigma_{0}^{\prime}$ by using Kramer-Kronig transformation~\cite{Bennett1965}.  Figs.~\ref{fig1}(c) and~\ref{fig1}(d) shows that $\sigma_{0}$ is robust against both the crystal chirality and N{\'e}el vector.
	
	In contrast, the off-diagonal elements of optical conductivity are substantially affected by the crystal chirality and N{\'e}el vector.  If $\bm{n}$ rotates within the (001) plane, two off-diagonal elements, $\sigma_{yz}$ and $\sigma_{zx}$, are nonzero.  The evolution of $\sigma_{yz}$ with $\bm{\chi}$ and $\bm{n}$ is plotted in Figs.~\ref{fig1}(e) and~\ref{fig1}(f), whereas the results of $\sigma_{zx}$ are shown in Figs. \textcolor{blue}{S1}(c) and \textcolor{blue}{S1}(d)~\cite{SuppMater}.  The signs of $\sigma_{yz}$ and $\sigma_{zx}$ are changed for $\bm{\chi}=+1$ and $-1$ states related by the symmetry $\mathcal{T}t_{1/2}$.  If $\bm{n}$ points along the [100] direction ($\theta = 90^{\circ}$ and $\varphi = 0^{\circ}$), $\sigma_{yz}$ is zero for both two chiral states due to the symmetry $\mathcal{M}_{[010]}t_{1/2}$.  By increasing $\varphi$ from $0^\circ$ to $90^\circ$, the magnitude of $\sigma_{yz}$ gradually increases for both two chiral states, which can be well accounted for the momentum-resolved optical Hall conductivity [Fig.~\ref{fig1}(g)].

\begin{figure}
		\centering
		\includegraphics[width=\columnwidth]{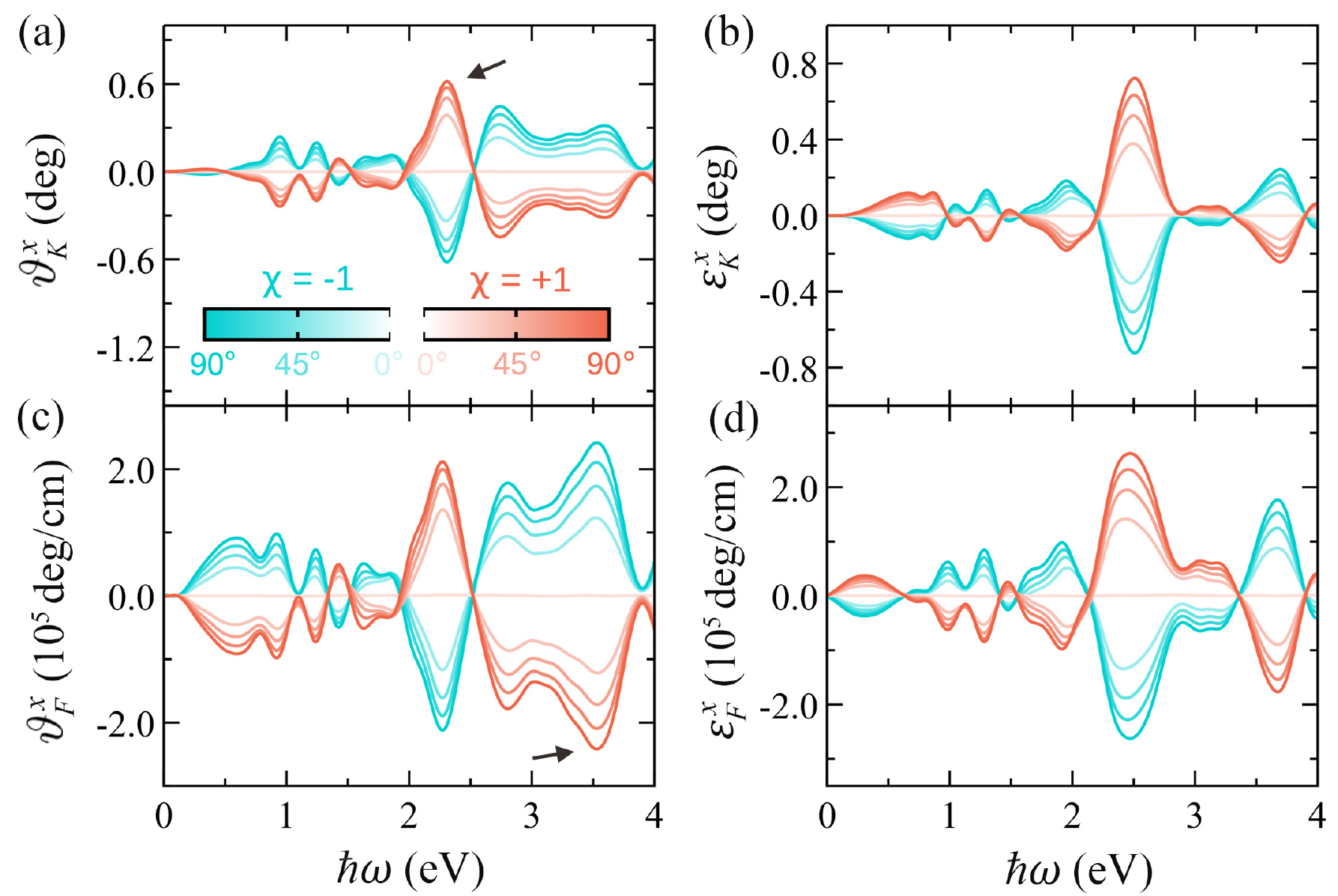}
		\caption{(a)(b) Kerr rotation angle and ellipticity and (c)(d) Faraday rotation angle and ellipticity for the left- and right-handed chirality states ($\bm{\chi}=\pm 1$) of RuO$_{2}$ when the N{\'e}el vector $\bm{n}$ rotates within the (001) plane ($\theta = 90^{\circ}$ and $0^\circ \leqslant \varphi \leqslant 90^\circ$).  The arrows in (a) and (c) mark the maximums of the Kerr and Faraday rotation angles at the photon energies of 2.31 and 3.53 eV, respectively.}
		\label{fig2}
	\end{figure}

	Now, we proceed to the CCMO effects for collinear AFM RuO$_{2}$, as depicted in Fig.~\ref{fig2}.  The Kerr and Faraday spectra ($\phi^{x}_{K}=\vartheta^{x}_{K}+i\varepsilon^{x}_{K}$ and $\phi^{x}_{F}=\vartheta^{x}_{F}+i\varepsilon^{x}_{F}$) exhibit a similar profile to the off-diagonal elements of optical conductivity, only differing by a minus~\cite{WX-Feng2017}.  The reason is simple as the off-diagonal elements of optical conductivity dominate the shape of MO spectra, while the diagonal elements mediate the amplitude of MO spectra (see Eqs. (\textcolor{blue}{S6}) and (\textcolor{blue}{S7}) in Ref.~\cite{SuppMater}).	 The reversal of the crystal chirality changes the signs of the Kerr and Faraday spectra but retains their magnitudes.  If $\varphi=0^\circ$, $\vartheta_{K,F}^{x}$ and $\varepsilon_{K,F}^{x}$ are zero due to the vanishing $\sigma^{x}\left(=\sigma_{yz}\right)$ [see Figs.~\ref{fig1}(e) and~\ref{fig1}(f)].  In the range of $0^{\circ} < \varphi \leqslant 90^{\circ}$, the magnitudes of $\vartheta_{K,F}^{x}$ and $\varepsilon_{K,F}^{x}$ increase monotonously with the increasing of $\varphi$.  An opposite trend appears to $\vartheta_{K,F}^{y}$ and $\varepsilon_{K,F}^{y}$ (see Fig. \textcolor{blue}{S2}).  It indicates that the magnitudes of CCMO effects can be effectively tuned by changing magnetization direction in collinear AFMs.  The largest Kerr and Faraday rotation angles of RuO$_{2}$ are 0.62 deg and 2.42$\times10^{5}$ deg/cm, respectively.  Particularly, the Kerr rotation angle is larger than that of traditional ferromagnets, e.g., bcc Fe ($\sim$0.6 deg)~\cite{GY-Guo1995}, hcp Co ($\sim$0.48 deg)~\cite{GY-Guo1994}, and fcc Ni  ($\sim$0.15 deg)~\cite{Engen1983}, and is also larger than that of famous noncollinear AFMs, e.g., Mn$_{3}X$ ($\sim$0.6 deg)~\cite{WX-Feng2015}, Mn$_{3}Y$ ($Y=$ Ge, Ga, Sn) ($\sim$0.02 deg)~\cite{Higo2018}, and Mn$_{3}Z$N ($\sim$0.42 deg)~\cite{XD-Zhou2019a}.  The large CCMO effects discovered in RuO$_{2}$ suggests a novel collinear antiferromagnetic platform for possible applications in MO recording beyond transitional ferromagnets~\cite{Mansuripur1995}.
	
\begin{figure*}
		\centering
		\includegraphics[width=2\columnwidth]{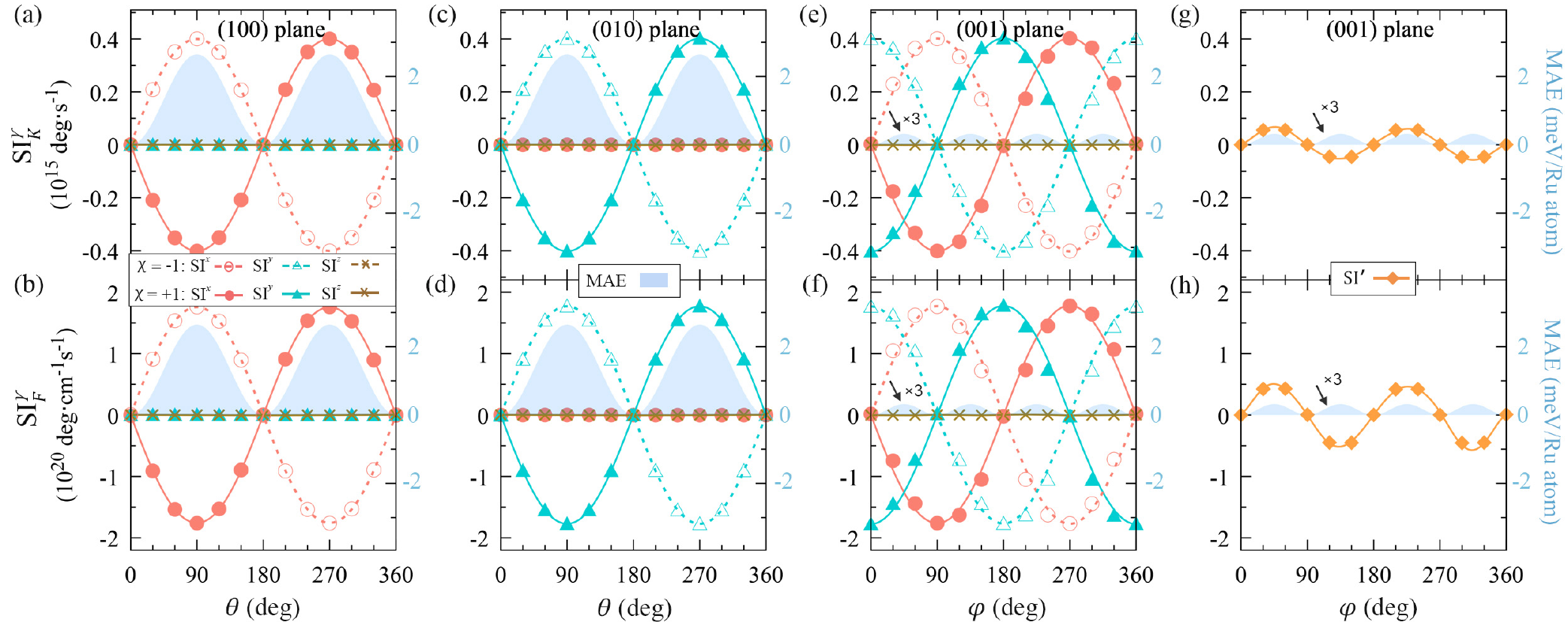}
		\caption{The SIs (colored symbols and lines, left axes) and MAE (shaded regions, right axes) for the left- and right-handed chirality states ($\bm{\chi}=\pm 1$) of RuO$_{2}$ when the N{\'e}el vector $\bm{n}$ rotates within the (100), (010), and (001) planes.  For the MAE on the (100)/(010) and (001) planes, the total energies when $\bm{n}$ points to the [001] and [100] directions are set to be the reference states, respectively.  The MAE on the (001) plane are multiplied by a factor of 3.  The SIs shown in (g) and (h) are subtracted by the ones when $\bm{n}$ points to the [100] direction.}
		\label{fig3}
	\end{figure*}

	Since the CCMO spectra are frequency-dependent quantities, additional information can be captured from their spectral integrals (SIs), defined as $\textnormal{SI}_{K}^\gamma=\int_{0^+}^{\infty}\vartheta_{K}^{\gamma}(\omega)\textnormal{d}\omega$ and $\textnormal{SI}_{F}^\gamma=\int_{0^+}^{\infty}\vartheta_{F}^{\gamma}(\omega)\textnormal{d}\omega$ with $\gamma=\{x,y,z\}$.  Figs.~\ref{fig3}(a) and~\ref{fig3}(b) show that when the N{\'e}el vector $\bm{n}$ rotates within the (100) plane, $\textnormal{SI}_{K}^{x}$ and $\textnormal{SI}_{F}^{x}$ are nonvanishing and their signs are opposite for two chiral states, which can be well understood from the symmetry requirements of $\vartheta_{K}^{x}$ and $\vartheta_{F}^{x}$.  Moreover, $\textnormal{SI}_{K}^{x}$ and $\textnormal{SI}_{F}^{x}$ exhibit a period of $2\pi$ with $\theta$, which is simply due to $\vartheta_{K}^{x}$ and $\vartheta_{F}^{x}$ are odd under time-reversal symmetry.  The MAE of two chiral states are equivalent and have a discrete two-fold degeneracy, $\textnormal{MAE}(\theta)=\textnormal{MAE}(\theta+\pi)$.  The most interesting finding is that the absolute values of SIs are proportional to the MAE, i.e., $|\textnormal{SI}_{K,F}^{x}|\propto\textnormal{MAE}$, which has not been reported in AFMs.  For the (010) plane, the nonvanishing $\textnormal{SI}_{K,F}^{y}$ display the same behaviors [Figs.~\ref{fig3}(c) and~\ref{fig3}(d)].  For the (001) plane, both $\textnormal{SI}_{K,F}^{x}$ and $\textnormal{SI}_{K,F}^{y}$ are nonvanishing and the MAE has a period of $\frac{\pi}{2}$ [Figs.~\ref{fig3}(e) and~\ref{fig3}(f)].  Nevertheless, the SIs are still proportional to the MAE if the SIs on the (001) plane are redefined as $\textnormal{SI}_{K}^\prime=\int_{0^+}^{\infty}\{[\vartheta_{K}^{x}(\omega)]^2+[\vartheta_{K}^{y}(\omega)]^2\}^{1/2}\textnormal{d}\omega$ and $\textnormal{SI}_{F}^\prime=\int_{0^+}^{\infty}\{[\vartheta_{F}^{x}(\omega)]^2+[\vartheta_{F}^{y}(\omega)]^2\}^{1/2}\textnormal{d}\omega$ [Figs.~\ref{fig3}(g) and~\ref{fig3}(h)].  Fig.~\ref{fig3} reveals that the SIs are closely related to the MAE for collinear AFMs, and the sign and size of SIs can be used to identify the crystal structure's chirality and the N{\'e}el vector's direction, respectively.
	
	\begin{figure}
		\centering
		\includegraphics[width=\columnwidth]{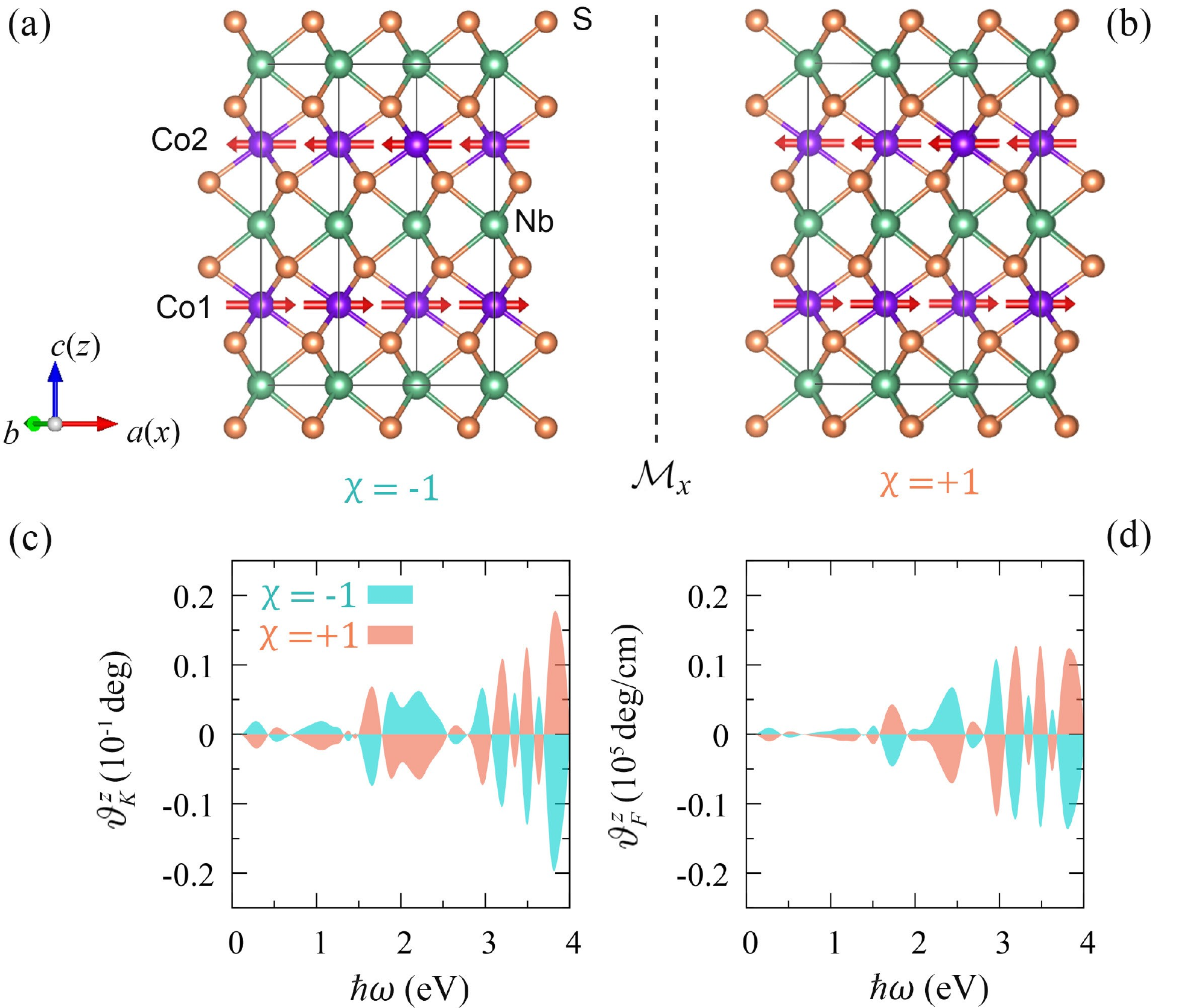}
		\caption{(a)(b) The crystal structures of CoNb$_{3}$S$_{6}$ with the right-handed ($\bm{\chi}=+1$) and left-handed ($\bm{\chi}=-1$) crystal chiralities.  The violet, green, and bronze spheres represent Co, Nb, and S atoms, respectively.  The red arrows label the spin orientations of magnetic Co atoms ($\bm{n}_{\textnormal{Co}1}$ and $\bm{n}_{\textnormal{Co}2}$).  The N{\'e}el vector, $\bm{n}=(\bm{n}_{\textnormal{Co}1}-\bm{n}_{\textnormal{Co}2})/2$, is along the [100] direction.  (c)(d) Kerr and Faraday rotation angles for the left- and right-handed chirality states of CoNb$_{3}$S$_{6}$.}
		\label{fig4}
	\end{figure}

	Finally, we demonstrate that the CCMO effects exist also in collinear AFMs with a \textit{global} crystal chirality, e.g., CoNb$_{3}$S$_{6}$.  The crystallographic space group of CoNb$_{3}$S$_{6}$ is $P6_{3}22$, which is compatible with chiral crystal structures due to the lacking of improper symmetry operations~\cite{Flack2003,Nespolo2018}.  As shown in Figs.~\ref{fig4}(a) and~\ref{fig4}(b), the left- and right-handed chiral structures are related to each other by a mirror plane $\mathcal{M}_{x}$, which preserves the positions of magnetic atoms and the orientations of spin magnetic moments but rearranges the nonmagnetic S atoms.  CoNb$_{3}$S$_{6}$ forms a collinear antiferromagnetic order below the N{\'e}el temperature of $\sim$25 K~\cite{Parkin1983,Ghimire2018,Tenasini2020}.  The magnetic space group of CoNb$_{3}$S$_{6}$ is $C2^\prime2^\prime2_{1}$ if two ferromagnetic Co layers magnetized along the [100] direction are antiferromagnetically coupled along the [001] direction via a NbS$_2$ layer.  The operation $C_{2z}t_{1/2z}$ ($t_{1/2z}=[0,0,0.5]$) forces the components of optical Hall conductivity that are perpendicular to the $C_{2}$ axis to be zero, and therefore gives rise to $\bm{\sigma}=[0,0,\sigma^{z}]$.  As a result, only the $z$ components of Kerr and Faraday rotation angles, $\vartheta_{K}^{z}$ and $\vartheta_{F}^{z}$, are nonvanishing, as shown in Figs.~\ref{fig4}(c) and~\ref{fig4}(d).  It is obvious that the MO effects for CoNb$_{3}$S$_{6}$ are chiral in the sense that the signs of $\vartheta_{K}^{z}$ and $\vartheta_{F}^{z}$ are reserved for the left- and right-handed crystal structures.  The reason is simple because the mirror symmetry $\mathcal{M}_{x}$ that relates two chiral crystal structures changes the signs of $\vartheta_{K}^{z}$ and $\vartheta_{F}^{z}$.  The CCMO effects uncovered in CoNb$_{3}$S$_{6}$ are characterized by the sign change resulting from the reversal of crystal chirality, similarly to the scenario of RuO$_{2}$.
	
	In summary, using the first-principles calculations together with group theory analyses, we have demonstrated the electrically controllable CCMO effects in collinear AFMs.  The sign and magnitude of CCMO spectra can be tuned by the crystal chirality and spin orientation, respectively.  Moreover, the spectral integrals are found to be proportional to the magnetocrystalline anisotropy energy.  The electrically controllable CCMO effects may provide a powerful probe of crystal chirality and spin orientation and be a promising MO recording technique based on antiferromagnetic materials.
	
	W.F., X.Z, and Y.Y. acknowledge the support from the National Key R\&D Program of China (Nos. 2020YFA0308800 and 2016YFA0300600), the National Natural Science Foundation of China (Nos. 11874085 and 11734003), and the Graduate Technological Innovation Project of Beijing Institute of Technology (No. 2019CX10018).  G.-Y. Guo acknowledges the support from the Ministry of Science and Technology, National Center for Theoretical Sciences, and the Far Eastern Y. Z. Hsu Science and Technology Memorial Foundation in Taiwan.
	

%

\end{document}